\documentclass[12pt]{article}

\usepackage[english]{babel}

\usepackage{latexsym}
\usepackage{amssymb}
\usepackage{epsfig}

\begin{document}

\title{Nonlinear electron dynamics in a rippled channel \\
       with time-dependent electric field: Quantum Arnol'd diffusion}

\author{V. Ya. Demikhovskii${}^{1}$, F.~M.~Izrailev${}^{2}$ and A. I. Malyshev${}^{1}$\\
{\it ${}^{1}$ Nizhny Novgorod State University, 603950,
} \\
{\it Nizhny Novgorod, Gagarin ave. 23, Russia} \\
{\it ${}^{2}$ Instituto de F\'{\i}sica, Universidad Aut\'{o}noma de Puebla,} \\
{\it Apdo. Postal J-48, Puebla, Pue. 72570, Mexico}}

\date{4th June 2006}

\maketitle

\begin{abstract}
We study the electron dynamics in a 2D waveguide bounded by a
periodically rippled surface in the presence of the time-periodic
electric field. The main attention is paid to a possibility of a
weak quantum diffusion along the coupling resonance, that can be
associated with the classical Arnol'd diffusion.  It was found
that quantum diffusion is possible only when the perturbation is
large enough in order to mix many near-separatrix levels.  The
rate of the quantum diffusion turns out to be less than the
corresponding classical one, thus indicating the influence of
quantum coherent effects. Another important effect is the
dynamical localization of the quantum diffusion, that may be
compared with the famous Anderson localization occurring in 1D
random potentials. Our estimates show that the quantum Arnol'd
diffusion can be observed in semi-metal rippled channels, for
which the scattering and decoherence times are larger than the
saturation time due to the dynamical localization.
\end{abstract}

\maketitle

\section{Introduction} 
In this paper we consider the quantum dynamics of a particle in
two-dimensional rippled channel subjected to time-dependent
electric field. In the absence of the electric field, the particle
motion can be either regular or weakly/strongly chaotic, depending
on the model parameters. In the case of regular motion, the linear
response to an external electromagnetic field can be described by
the famous Kubo formula. In the opposite case of a strong chaos
one can use the approach recently developed in
Refs.~\cite{WA92,BSK03}. The situation with weak chaos is quite
peculiar and needs a special approach. Of specific interest is the
case when chaos is extremely weak, and occurs in narrow regions of
slightly destroyed non-linear resonances, thus leading to the
so-called Arnol'd diffusion {\it along} the resonances
\cite{C79,LL92}. Our further analysis enlightens the connection
between quantum and classical mechanisms of the Arnol'd diffusion
on the rippled channel model. The results presented here are based
on the theory developed in Ref.~\cite{DIM02}, and expand those
preliminary reported in Ref.~\cite{DIM06}.

The dynamical chaos in classical Hamiltonian systems is related to
the destruction of separatrices of nonlinear
resonances~\cite{C59}. For a weak interaction, chaotic motion
occurs only in the vicinity of separatrices of the resonances. On
the other hand, inside the resonances the motion remains regular,
in accordance with the Kolmogorov-Arnol'd-Moser (KAM) theory (see,
for example, Ref.~\cite{LL92}). If the number $N$ of degrees of
freedom larger than $2$, the KAM surfaces do not separate the
stochastic layers, therefore, they form a stochastic web that can
cover whole phase space of a system. Thus, if trajectory starts
{\it inside} the stochastic web, it can diffuse throughout the
phase space. Such a diffusion along stochastic webs was predicted
by Arnol'd in 1964~\cite{A64}, and since that time it is known as
an universal mechanism for instability and chaos in generic
nonlinear Hamiltonian systems with $N > 2$ (see, for example,
review~\cite{C79} and references therein).

The chaotic dynamics of a quantum systems under time dependent
periodic per\-tur\-ba\-tion was studied in a series of works. The
authors of Ref.~\cite{WA92} in the frame of random matrix theory
have investigated the spectral properties of evolution operator in
the generic system with a time-dependent Hamiltonian. The
statistical properties of quasienergy spectrum, the localization
of the eigenstates of evolution operators as well as the process
of saturation of the energy absorption in the external periodic
field was considered. In the paper~\cite{BSK03} the response of
the quantum dot electron system to a periodic perturbation was
studied analytically in terms of zero-dimensional time-dependent
nonlinear sigma model. The quantum correction to the energy
absorption rate as a function of the dephasing time was
calculated. In particular, it was shown that the dynamical
localization corrections similar to the $d$-dimensional weak
localization corrections to conductivity if the perturbation is a
sum of $d$ incommensurate harmonic functions. A typical
application of these results would be to the response of a
dif\-fe\-rent elec\-tron me\-so\-sco\-pic systems (quantum wells,
wires and dots) to electromagnetic radiation.
Experimentally the response of
2D electron gas in quantum dot formed in GaAs/AlGaAs
heterojunction to electromagnetic radiation was investigated
in~\cite{DMH03,H99}, where the effect of absorption saturation in
open chaotic quantum dots was observed in particular.

The chaotic nature of the free particle dynamics in a rippled
channel (without external electromagnetic fields) has been
investigated in Refs.~\cite{LKRH96,LNRK96,LMI01,IML03}, both in
classical and quantum models. In particular, in Ref.~\cite{LKRH96}
the transport properties were considered in a ballistic regime.
The energy band structure, eigenfunctions and density of states
have been analyzed in Refs.~\cite{LNRK96,LMI01}. The structure of
quantum states in the channel with rough boundaries, including the
phenomena of quantum localization, have been studied in
Ref.~\cite{IML03}. In particular, it was found that the eigenstates
are very different in their localization properties.

The paper is organized as follows. In Sec.~2 we discuss the
classical model and the mechanism of the Arnol'd diffusion in the
rippled channel. Stationary quantum states corresponding to the
coupling resonance are studied in Sec.~3. In Sec.~4 we build the
evolution operator for one period of the external field, and
discuss its structure. In the same section both classical and
quantum diffusion coefficients are calculated as a function of the
goffer amplitude. The nature of the quantum diffusion suppression,
as well as the dynamical localization, are also discussed here. In
last section we formulate main results obtained in the paper, and
discuss the parameters for which the quantum Arnol'd diffusion
could be observed experimentally.

\section{Classical Arnol'd diffusion in 2D rippled channel} 
\begin{figure}[!tb]
 \begin{center}
  \includegraphics[width=4.5in]{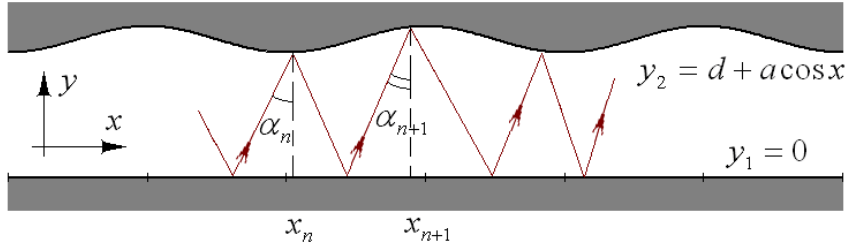}
 \end{center}
 \caption{An example of the electron trajectory in the rippled channel.}
 \label{fchan}
\end{figure}

We study the Arnol'd diffusion in a periodic two-dimensional
waveguide, see Fig.~\ref{fchan}. It is defined by the upper
profile given in dimensionless variables by the function
$y=d+a\cos{x}$ and by the low profile that is assumed to be flat,
$y=0$. Here $d$ is the average width, $a$ is the ripple amplitude,
and the goffer period is equal to $2\pi$. In what follows we put
the ratio $a/d$ is small, in order to avoid the global chaos that
occurs for $a/d \sim 1$. The collisions of a particle with these
boundaries are assumed to be elastic.

The Poincar{\'e} map for the positions and angles $(x_n,\alpha_n)$
of the reflections from the upper wall is determined by
  \begin{eqnarray}
   \label{map}
    \cases{
    \alpha_{n+1}=\alpha_n-2\arctan(a\sin{x_n}), \cr
    x_{n+1}=x_n+\tan{\alpha_{n+1}} \bigl(2d
     +a(\cos{x_n}+\cos{x_{n+1}})\bigr),}
  \end{eqnarray}
Here $x_n$ is the position corresponding
to the {\it n}-th bounce at the rippled wall, and $\alpha_n$ is
the angle of the particle trajectory makes with the vertical at
$x=x_n$.

\begin{figure}[tb]
\begin{center}
\includegraphics[width=5.5in]{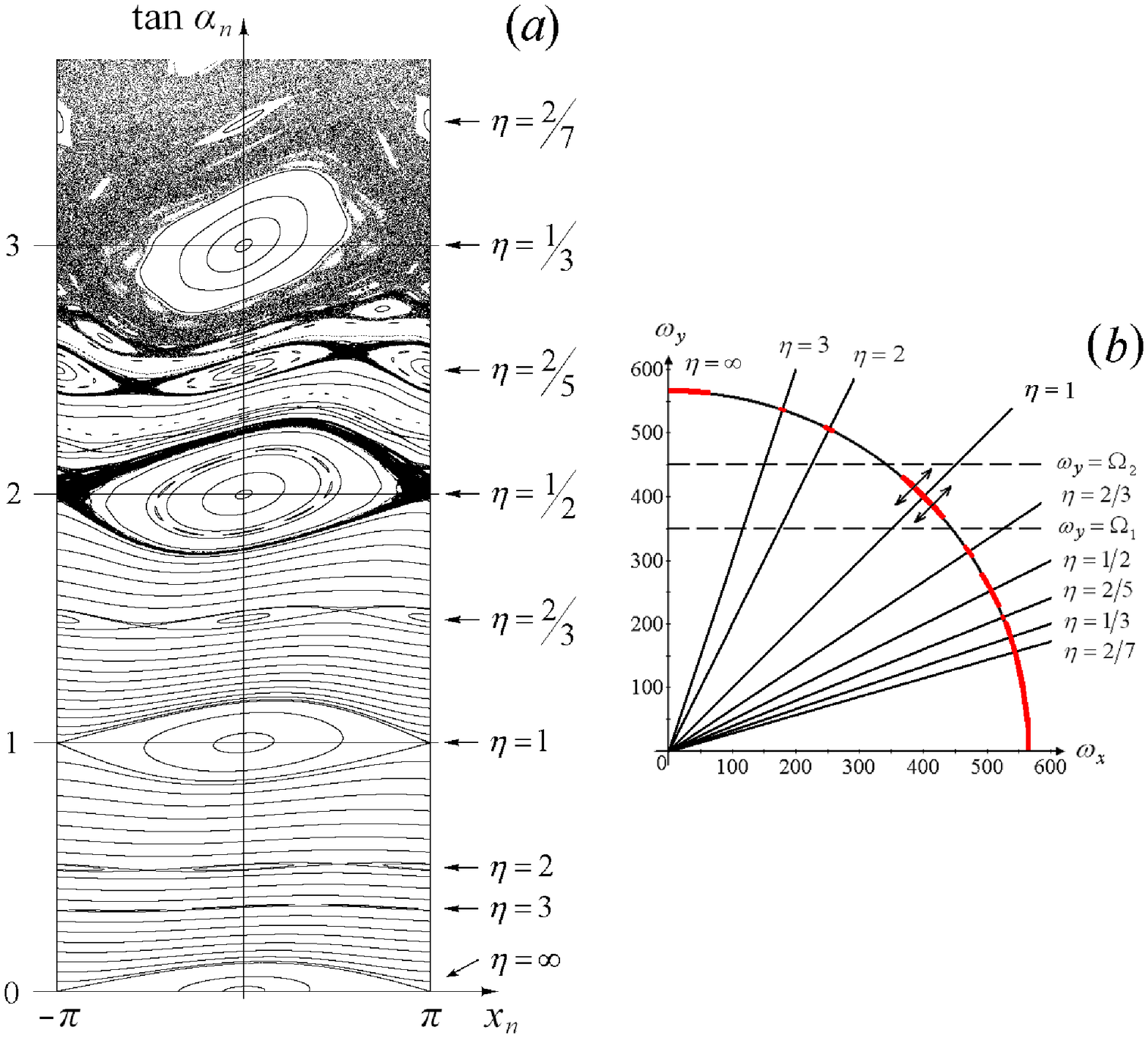}
\caption{The resonance structure for the rippled channel model.
({\it a}) The Poincar{\'e} section is shown for $d=\pi$ and
$a=0.01$, demonstrating the structure of some of coupling
resonances. ({\it b}) The positions of some coupling resonances
and two driving resonances (dashed lines)
and the isoenergetic curve $E=1.6\cdot 10^5$ on the
frequency plane are shown, together with the resonances widths
(by red color).}
\label{fclass}
\end{center}
\end{figure}

An example of the trajectory described by this map is shown in
Fig.~\ref{fclass}({\it a}). It can be shown that there are many
nonlinear resonances arising in this model due to the coupling
between two degrees of freedom, with the resonance condition,
$\eta =T_x/T_y =\omega_y /\omega_x$. Here $T_y$ is the period of a
transverse oscillation inside the channel, $T_x$ is the time of
flight of a particle over one period of the waveguide, $\omega_x$
and $\omega_y$ are the corresponding frequencies, and $\eta$ is
the rational number. It should be noted that in the neighborhood
of the resonances~$\eta=1/n$ (for which $n=0, 1, 2,\dots$) it is
possible to reduce the map~(\ref{map}) to the well known standard
map~\cite{C79} with parameter $K_n=4ad(1+(\pi n/d)^2)$.

The mechanism of the classical Arnol'd diffusion in this system is
illustrated in Fig.~\ref{fclass}({\it b}). Here some of the
resonance lines for the different values of $\eta$ shown in
Fig.~\ref{fclass}({\it a}), are presented with the use of the
$\omega_x$, $\omega_y$-plane. The curve of a constant kinetic
energy~$E$ is also shown here, determined by the equation
\begin{eqnarray}
  \label{res_omega}
  \omega_x^2 + \left(\frac{\omega_y d}{\pi}\right)^2 = \frac{2E}{m},
\end{eqnarray}
where $m$ is the particle mass that we will set to unity.

As one can see, there are two different types of resonances. The
resonances with $\eta\le 1/3$ are overlapped and in this region
the global chaos regime is realized. All other resonances are not
overlapped, and surrounded by narrow near-separatrix stochastic
regions. These resonances are isolated one from another by the
KAM-surfaces, and therefore, for a weak perturbation the
transition between their stochastic layers is forbidden. Such a
transition could occur in the case of the resonance overlap only.
In the absence of an external field, the passage of a trajectory
{\it along} any stochastic layer (this direction is shown in
Fig.~\ref{fclass}({\it b}) for resonance $\eta=1$ by two arrows)
is also impossible because of the energy conservation. The external
time-periodic field removes the latter restriction, and slow diffusion
along stochastic layers becomes possible. As a result, the particle
remains located on the coupling resonance, however, with a
proportional change of its momentum components.

The external electric field is given by the potential
$V(y,t)=-f_0y(\cos \Omega_1 t + \cos \Omega_2 t)$, giving rise to
two main resonances, $\omega_y = \Omega_1$ and
$\omega_y=\Omega_2$. In order to calculate the diffusion rate, we
consider a part of the Arnol'd stochastic web created by three
resonances, namely, by the coupling resonance $\omega_x=\omega_y$
and two driving resonances with frequencies $\Omega_1$ and
$\Omega_2$. Correspondingly, we choose the initial conditions
inside the stochastic layer of the coupling resonance. To avoid a
strong overlap of the resonances, however, to provide a weak chaos
in the near-separatrix region, we assume that the relation
$a/f_0=10^{-3} \ll 1$ is fulfilled.

\section{Stationary states at coupling resonance} 
In the following it is convenient to rewrite the problem in the
curvilinear coordinates $\tilde x^i$ for which the both channel
boundaries appear to be flat~\cite{DPS83}. As a result, the
covariant coordinate representation of the Schr\"odinger equation
has the form
\begin{eqnarray}
   \label{schr}
  -\frac{1}{2 \sqrt {g}} \frac {\partial}{\partial
   \tilde x^i} \sqrt {g} g^{ij} \frac {\partial \psi}{\partial \tilde x^j}
   =E \psi
\end{eqnarray}
where $g^{ij}$ is the metric tensor and $g\equiv \det (g_{ij})$.
Here we use the units in which the Plank's constant and effective
mass are equal to unity. Let the new coordinates are given by the
relations,
\begin{eqnarray}
  \label{transf} \tilde x=x, \,\,\,\,\,\,\,\,\,\,\, \tilde y=\frac{y}{1+\epsilon\cos x}
\end{eqnarray}
where $\epsilon=a/d$. Then the boundary conditions
read as $\psi(\tilde x,0)=\psi(\tilde x,d)=0$ and the metric
tensor is
\begin{eqnarray}
  g^{ij}= \left(
             \matrix{
             1 \,\,\,& \frac{\epsilon \tilde x \sin \tilde x}{1+\epsilon \cos \tilde x} \cr
             \frac{\epsilon \tilde x \sin \tilde x}{1+\epsilon \cos \tilde x}
             \,\,\,&
             \frac{1+\epsilon ^2 \tilde x^2 \sin ^2 \tilde x}{(1+\epsilon \cos \tilde x)^2}
                    }
         \right),
  \label{tensor}
\end{eqnarray}
with the orthonormality condition,
\begin{eqnarray}
   \label{ortho}
   \int \psi_i^{\star} \psi_j\sqrt{g}d\tilde x d\tilde y=\delta_{ij}.
\end{eqnarray}
If the ripple amplitude $a$ is small compared to the channel width
$d$, one can safely keep only the first-order terms in $\epsilon$
in the Schr\"odinger equation~(\ref{schr}). This strongly
simplifies numerical simulations without the loss of generality.
However, one should note that with an increase of roughness in the
scattering profile $y(x)$ this approximation may be invalid, due
to an influence of the so-called gradient scattering, see details
in Refs.\cite{IML03}. As a result, we obtain the following
Hamiltonian~\cite{DPS83},
\begin{eqnarray}
  \label{ham}
\hat H = \hat H_0 + \epsilon \hat U
  = - \frac{1}{2} \left(\frac{\partial^2}{\partial x^2}
  + \frac{\partial^2}{\partial y^2}\right)  \cr
  + \frac {\epsilon}{2}
  \left( 2\cos x  \frac{\partial^2}{\partial y^2} -2 y \sin x
  \frac{\partial^2}{\partial x \partial y} - y \cos x
  \frac{\partial}{\partial y} - \frac{1}{2} \cos x - \sin x
  \frac{\partial }{\partial x} \right).
\end{eqnarray}
\noindent Here and below we omitted tildes in coordinates
$\tilde x$ and $\tilde y$.

Since the Hamiltonian is periodic in the longitudinal coordinate
$x$, the eigenstates are Bloch states characterized by the Bloch
index $k$. Therefore, the eigenstates $\psi^k (x,y)$ of the total
Hamiltonian $\hat H$ can be presented as follows,
\begin{eqnarray} \psi^k (x,y) =
  \sum_{n,m} c^k_{nm} \psi^0_{nm}(k,x,y)
  \label{psi}
\end{eqnarray}
where
\begin{eqnarray}
   \label{psi0}
   \psi^0_{nm}(k,x,y)=\frac{1}{\sqrt{\pi d}}
   e^{i(n+k)x}\sin \left( \frac {\pi my}{d}\right)
\end{eqnarray}
are the eigenstates of the unperturbed Hamiltonian $\hat H_0$.
Here we normalize wave functions to the length $L=2\pi$ of the
period in $x$-direction. In the absence of the perturbation
($\epsilon=0$) the energy spectrum has the form
\begin{eqnarray}
E^0_{nm}(k) = \frac{1}{2} \left( (n+k)^2 + \frac{\pi ^2 m^2}{d^2}
\right ).
\end{eqnarray}
The Bloch wave vector $k$ has continuous values, in particulary,
$-1/2 \leq k \leq 1/2$ in the first Brillouin zone.

Now we proceed to solve the system of algebraic equations for the
coefficients $c^k_{nm}$,
\begin{eqnarray}
  E(k)c^k_{nm} = E^0_{nm}(k) c^k_{nm}
  + \epsilon \sum_{n',m'} U^k_{nm,n'm'} c^k_{n'm'},
  \label{system}
\end{eqnarray}
with $-\infty < n < \infty$ and $m=1, ... , \infty$.
The matrix elements in~(\ref{system}) are defined as
\begin{eqnarray}
  \label{matel}
 U^k_{nm, n'm'}=\int \left( \psi^0_{(k+n'),m'}
  \right)^{\star} \hat U(x,y) \psi^0_{(k+n),m} dx dy
  = - \frac{1}{2} \Biggl[ \frac{\pi^2 m^2}{d^2} \left(
  \delta_{n',n+1} + \delta_{n',n-1} \right) \delta_{m,m'}   \cr
   + \frac{\left(-1\right)^{m+m'}mm'}{m^2-m'^{2}}
   \Bigl( \bigl( 1+2(k+n)\bigr)
  \delta_{n',n+1} + \bigl( 1-2(k+n)\bigr) \delta_{n',n-1} \Bigr)
  \Biggr].
\end{eqnarray}

Following Refs.~\cite{DIM02}, we analyze the energy spectrum in
the vicinity of the main coupling resonance $\eta=1$ determined by
the condition $\omega_{n_0}=\omega_{m_0}$ with $\omega_{n_0} =
E_{n_0+1}(k)- E_{n_0}(k)= k + n_0 +1/2$ and
$\omega_{m_0}=E_{m_0+1}- E_{m_0}=\pi^2(2m_0+1)/2d^2$. In a deep
semiclassical region for $n_0 \gg 1$ and $m_0 \gg 1$, one can
write $\omega_{n_0} \approx n_0$ and $\omega_{m_0} \approx \pi^2
m_0/d^2$. It should be noted that the similar resonance condition
can be satisfied for negative $n_0$ as well, when
$-n_0\approx\pi^2 m_0/d^2$, that corresponds to the motion in the
opposite direction. Since below we assume large values, $|n|\gg
1$, one can neglect the tunneling from the resonances with $n>0$
to those with $n<0$. Note also that for the values of $k$ at the center
of the energy band, $k=0$, as well as at the band edges, $k= \pm
1/2$, there are additional integrals of motion due to anti-unitary
symmetry. For this reason in what follows we consider generic case
of other values of $k$. The properties of the energy spectra and
eigenstates for specific values $k=\pm 1/2$ and $k=0$ will be
discussed separately.

In the vicinity of the coupling resonance it is convenient to
introduce new indexes $r=n-n_0$ and $p=r+(m-m_0)$. In this notation
instead of the system~(\ref{system}) we obtain,
\begin{eqnarray}
   \label{systemlast}
   E(k) c^k_{rp}
   = \left( p \omega_{m_0} + \frac{r^2}{2} +
   \frac{\pi^2}{2d^2}\left( p-r \right)^2 +kr \right) c^k_{rp}
   + \sum_{r',p'} U^k_{r p, r' p'} c^k_{r'p'},
\end{eqnarray}
\noindent where energy $E(k)$ counts from the level $E^0_{n_0m_0}(k)$.

With equation~(\ref{systemlast}) one can investigate quantum
states at the coupling resonance for different values of the wave
number~$k$. For example, by considering the states near some
positive value $n_0$ and for $-1/2 < k < 1/2$ (with $k\ne 0$), we
obtain that the energy spectrum consists of series of Mathieu-like
groups. These groups are separated one from another by the energy
$\omega_{n_0}$. The structure of energy spectrum in each group is
typical for any quantum nonlinear resonance. Namely, the lowest
levels are practically equidistant, and the accumulation point
corresponds to the classical separatrix. However, in contrast with
the well known Mathieu spectrum, all the above-separatrix states
are non-degenerate even for the zero goffer $a = 0$.

Besides the above discussed part of the spectrum, another series
of groups occurs in the same energy region, that corresponds to
the similar resonance with the negative values of $n_0$. The
eigenstates corresponding to this resonance are the waves
propagating in the negative direction of the $x$~axis.

\begin{figure}[!tb]
\begin{center}
\includegraphics[width=3.5in]{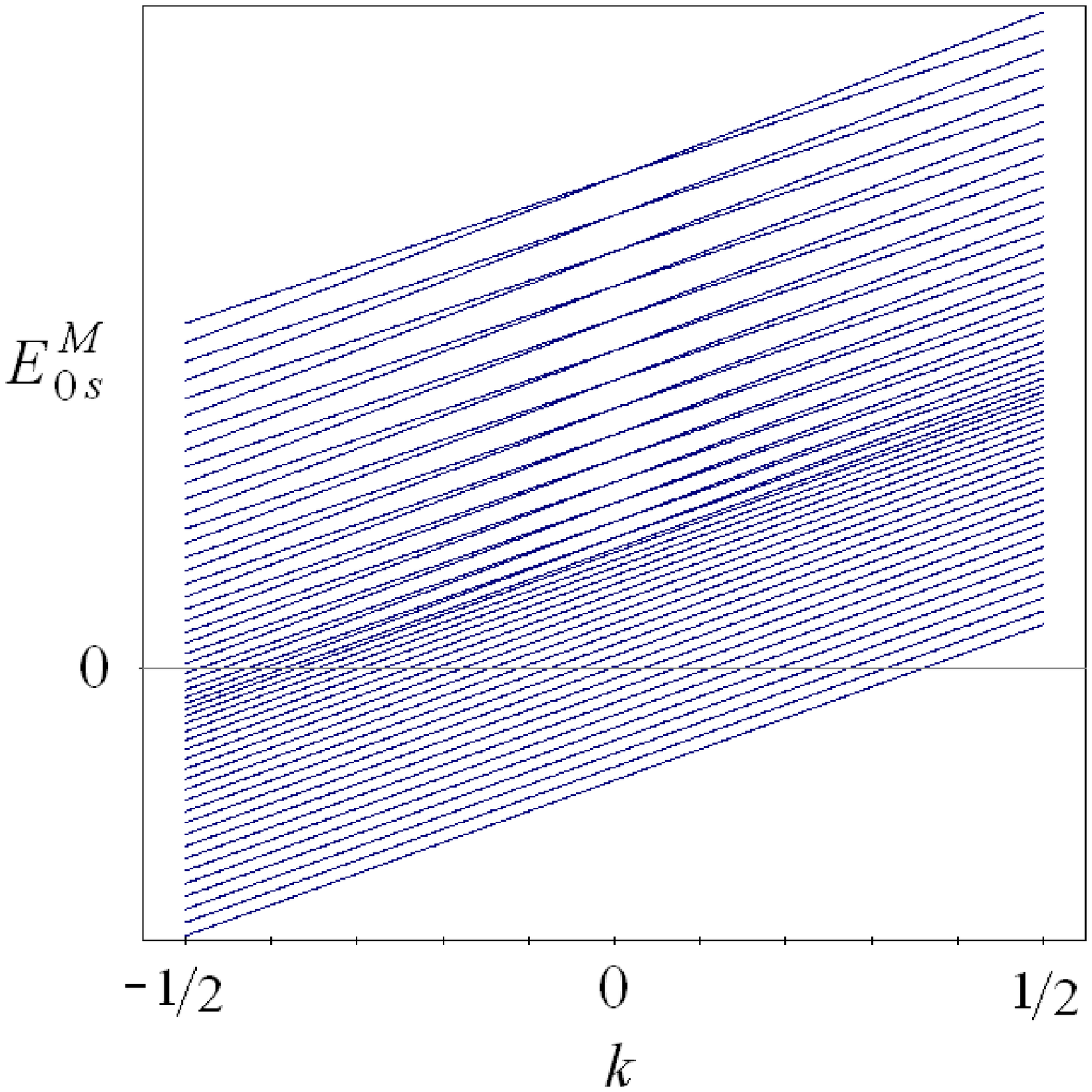}
\caption{Structure of the energy spectrum in dependence on the
Bloch number $k$, for $q=0$ and $n_0=m_0=400$, $d=\pi$ and
$a=0.003$, see text.}
\label{E_k}
\end{center}
\end{figure}

In accordance with the spectrum structure it is convenient to
characterize the states at the coupling resonance by two indexes:
the group number~$q$, and level number~$s$ characterizing the
levels inside the group. Correspondingly, the energy of each group
can be written in the form,
\begin{eqnarray}
  \label{energy}
  E_{q,s}(k) =\omega_{n_0}(k) q + E_{q,s}^{M}(k),
\end{eqnarray}
where $E_{q,s}^{M}(k)$ is the Mathieu-like spectrum for one group.
The indexes $q$ and $s$ correspond to fast and slow variables
characterizing the motion inside the classical coupling resonance.
Fig.~\ref{E_k} illustrates the structure of the spectrum for one
group with $q=0$, in dependence on $k$. The states corresponding
to $n_0>0$ shown in the figure, have positive values of the
derivative $\partial E/\partial k$.

Let us discuss now the resonance states for specific values
$k =0$ and $k=\pm 1/2$. The corresponding states have no classical
counterparts because the waves propagating in opposite directions
are coupled due to the Bragg condition. Indeed, at the center and
at the edges of the Brillouin zone, where the group velocity
vanishes, $\partial E/\partial k=0$, the Fourier series
(\ref{psi}) of the Bloch function $\psi^k(x,y)$ contains the terms
with positive and negative~$n$ equally. As a result, the Bloch
function $\psi^k(x,y)$ is the even (and real) or odd (and pure
imaginary) standing wave. In contrast, the Bloch functions for
other values of $k$ have no definite parity and can be
characterized by a positive or negative momentum.

As one can see, in Eqs.~(\ref{psi}) and (\ref{psi0}) for $k=0$ it
is more convenient to use the functions $\cos nx$ and $\sin nx$,
instead of the exponents $e^{inx}$. Then, the matrix elements
$U^k_{nm, n'm'}$ are not different from the analogous
ones~(\ref{matel}), and, therefore, the spectrum structure is
similar to that discussed before for generic values of $k$. In a
similar way, the functions $\cos (n\pm 1/2)x$ and $\sin(n\pm
1/2)x$ should be used in the expansion~(\ref{psi}) for $k=\pm
1/2$.

The resonance states for $k = 0$ and $k=\pm 1/2$ are either even
or odd, therefore, the energy spectrum consists of two series of
Mathieu-like groups corresponding to a specific parity. For $k=0$
the above-separatrix levels in each group are practically non-
degenerated, in contrast with the case of generic values of $k$,
see Fig.~\ref{E_k}.

\begin{figure}[!tb]
 \begin{center}
 \includegraphics[width=6.0in]{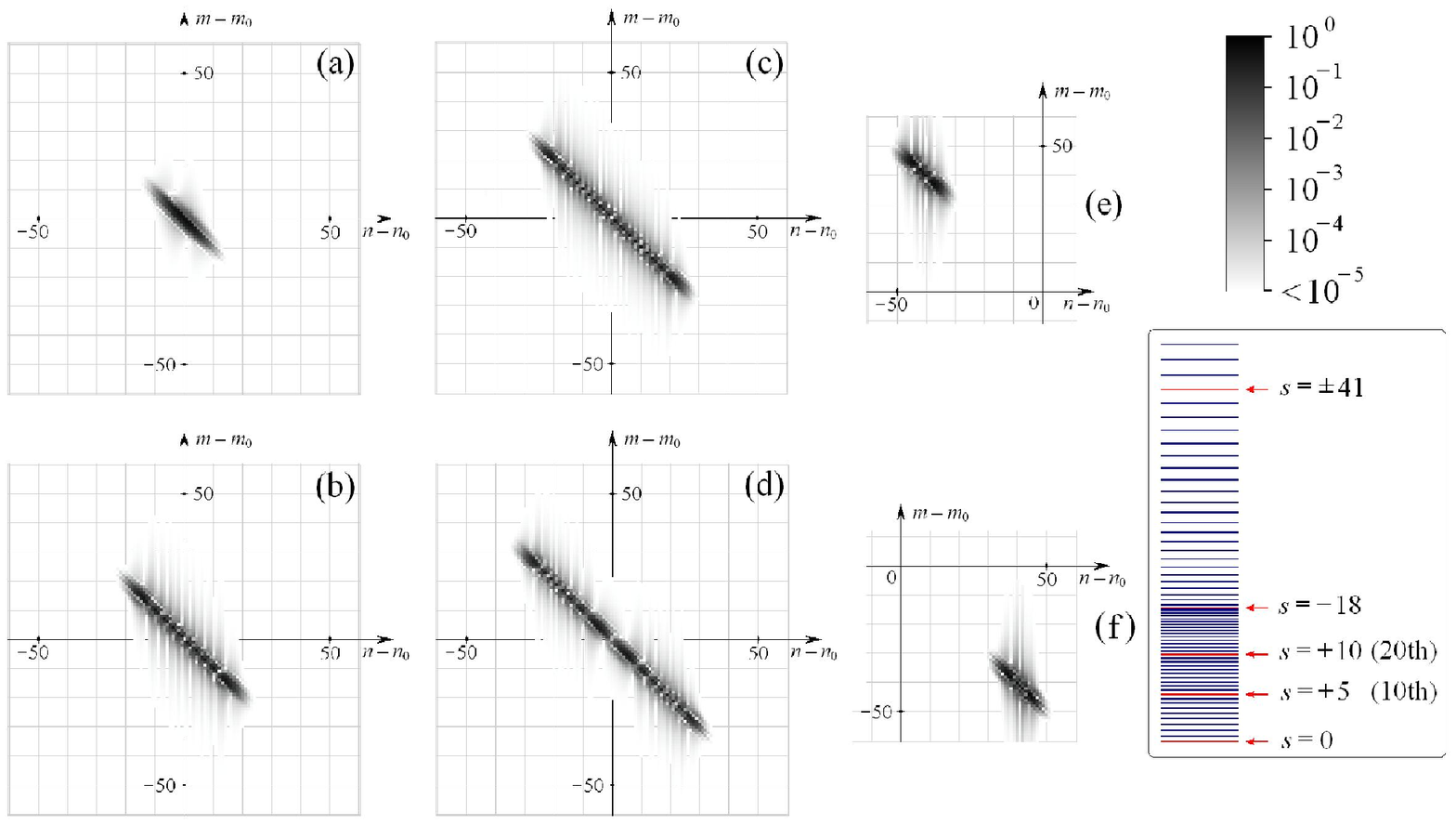}
\caption{Probability $|c^k_{rp}|^2$ in grayscale for different
resonant quantum states ($\eta=1$) corresponding to the
Mathieu-like group $q = 0$ (see inset), for parameters $d=\pi$,
$a = 0.005$ and $k=0.01$, and for $n_0=m_0=400$.
 ({\it a}) lowest level (ground state) $s = 0$;
 ({\it b}) 10th state ($s = +5$); ({\it c}) 20th state ($s = +10$);
 ({\it d}) near-separatrix level $s = -18$;
 ({\it e}), ({\it f}) above-separatrix levels $s=\pm 41$.}
 \label{distr1}
 \end{center}
\end{figure}

\begin{figure}[!tb]
 \begin{center}
 \includegraphics[width=6.0in]{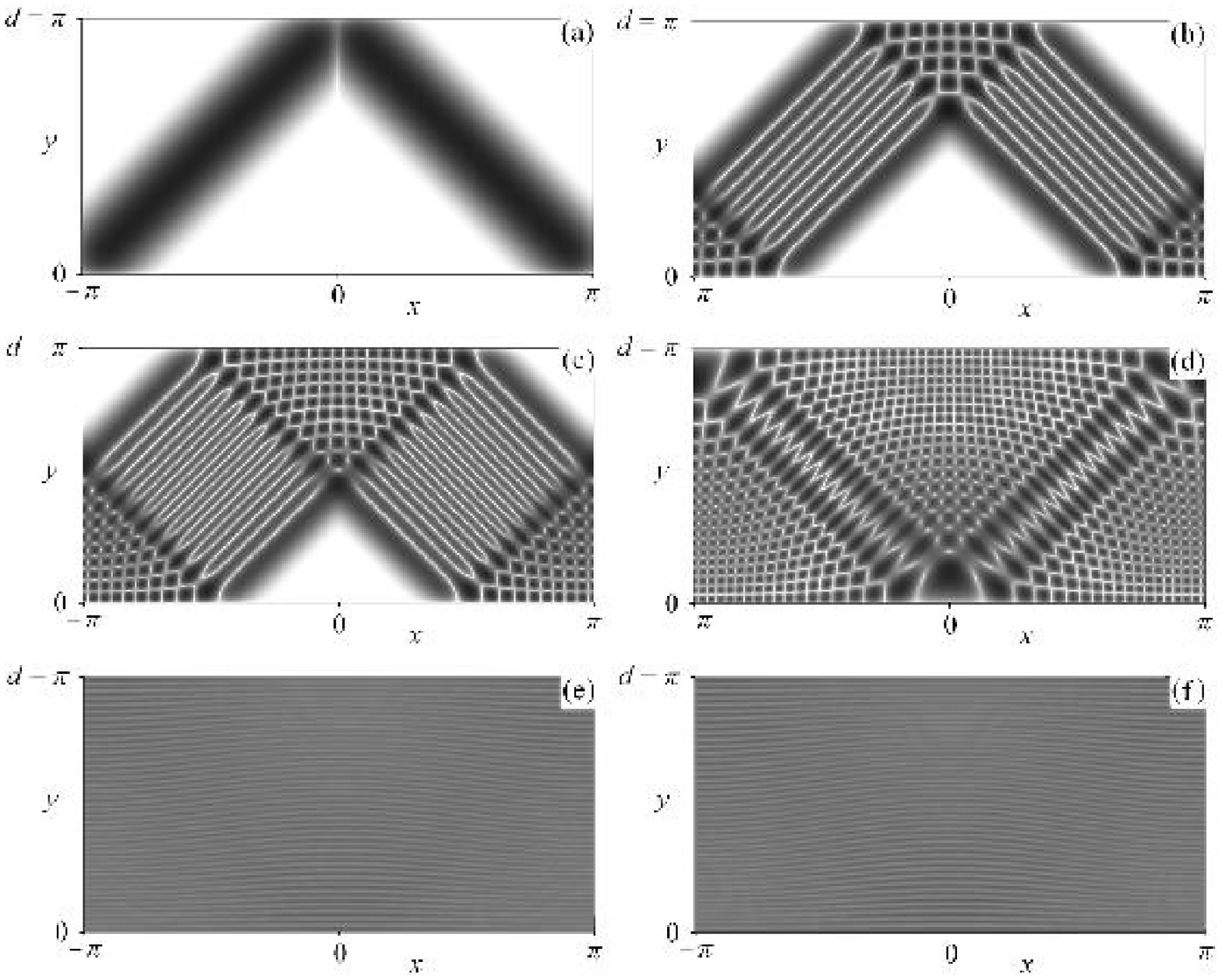}
 \caption{Examples of eigenstates probabilities
 $|\psi^k_{q,s}(x,y)|^2$ in the grayscale for the same model
 parameters and quantum state numbers as in Fig.~\ref{distr1}.}
 \label{distr2}
 \end{center}
\end{figure}

The wave functions of the resonance quantum states have very rich
and complex structure. To make it clear
we plotted the probability distribution
of the eigenvectors of the Eq.~(\ref{systemlast})
with different $q$ and $s$ in the unperturbed basis~(\ref{psi0})
of the Hamiltonian $\hat H_0$, see Fig.~\ref{distr1}.
The probability distribution $|\psi^k_{q,s}(x,y)|^2$
in $(x,y)$-space for the same states are shown at Fig.~\ref{distr2}.
First, let us discuss the
results shown in Fig.~\ref{distr1}. Here and below, the ground
states in each group have the number $s=0$, and all other states
are reordered according to the energy increase, labelling by $1,
-1, 2, -2, \dots$ etc.

So, in Fig.~\ref{distr1}(a) one can see
probability distribution~$|c^k_{rp}|^2$, corresponding
to the lowest level in the group with $q=0$.
In Figs.~\ref{distr1}(b,c) the probability distributions
corresponding to the 10th and 20th resonance levels of the same
group are depicted. To compare with, we show in
Fig.~\ref{distr1}(d) the distribution corresponding to the level
$s=-18$ taken from the near-separatrix region. All these states
are symmetric with respect to $n-n_0=0$ and $m-m_0=0$. It is seen
that the degree of delocalization of these eigenfunctions in the
diagonal direction increases with an increase of the energy.
Specifically, the separatrix state is much more extended in the
unperturbed basis, in comparison with the state at the resonance
center. As will be shown below, the separatrix resonance states,
having a maximal variance in the unperturbed basis, provide the
Arnol'd diffusion. The eigenfunctions corresponding to two
near-degenerate levels above the separatrix (with $s=\pm 41$),
are shown in Fig.~\ref{distr1}(e,f). In contrast,
the probability distributions corresponding
to above-separatrix states are non-symmetric (their
maximums are shifted with respect to the origin), and have a
relatively small variance.

Fig.~\ref{distr2} illustrates the probability distribution for
same resonance eigenstates in the transformed
$(x, y)$-coordinates. Note that for small values $a/d \ll 1$ the
probability distribution in original coordinates has a practically
similar shape. It is clear that the structure of these states
is quite different from the unperturbed ones. A regular pattern in
Fig.~\ref{distr2}(a), plotted for the state $s = 0$, resembles the
classical trajectory corresponding to the resonance center. This
orbit starts at point $x = -\pi$, $y=0$ and ends at $x = \pi$,
$y=0$. The reflection point is located at $x=0$, $y=\pi$ where the
width of the channel has a maximum. In accordance with the
uncertainty principle, this distribution has a nonzero width.
The probability distributions in Figs.~\ref{distr2}(b,c),
corresponding to the 10th and 20th resonance states, also resemble
near-resonance classical trajectories, however, have a complex
internal structure. Each state of this kind corresponds to the
group of classical trajectories in $(x, y)$-space, for which
initial conditions lay at one closed orbit in the phase space
($\sin\alpha$, $x$). One can see that the number of white
longitudinal lines (minima) are equal to the number of resonance
state minus one. Fig.~\ref{distr2}(d) illustrates the probability
distribution for one of the near-separatrix states, $s=-18$. Here
the pattern also originated from classical unstable trajectory. It
should be stressed that for all presented states the probability
distributions mainly have a non-chaotic character. In
Fig.~\ref{distr2}(e,f) we demonstrate the probability
distributions for above-separatrix states $s=\pm 41$. The
distributions for these states are uniform, indicating that they
are weakly perturbed. In conclusion, we would like to point out
that the electron density distribution shown in Figs.~\ref{distr2}
resemble the electromagnetic field distribution in quasi-optic
regimes~\cite{G98}.

\begin{figure}[!tb]
 \begin{center}
 \includegraphics[width=6.0in]{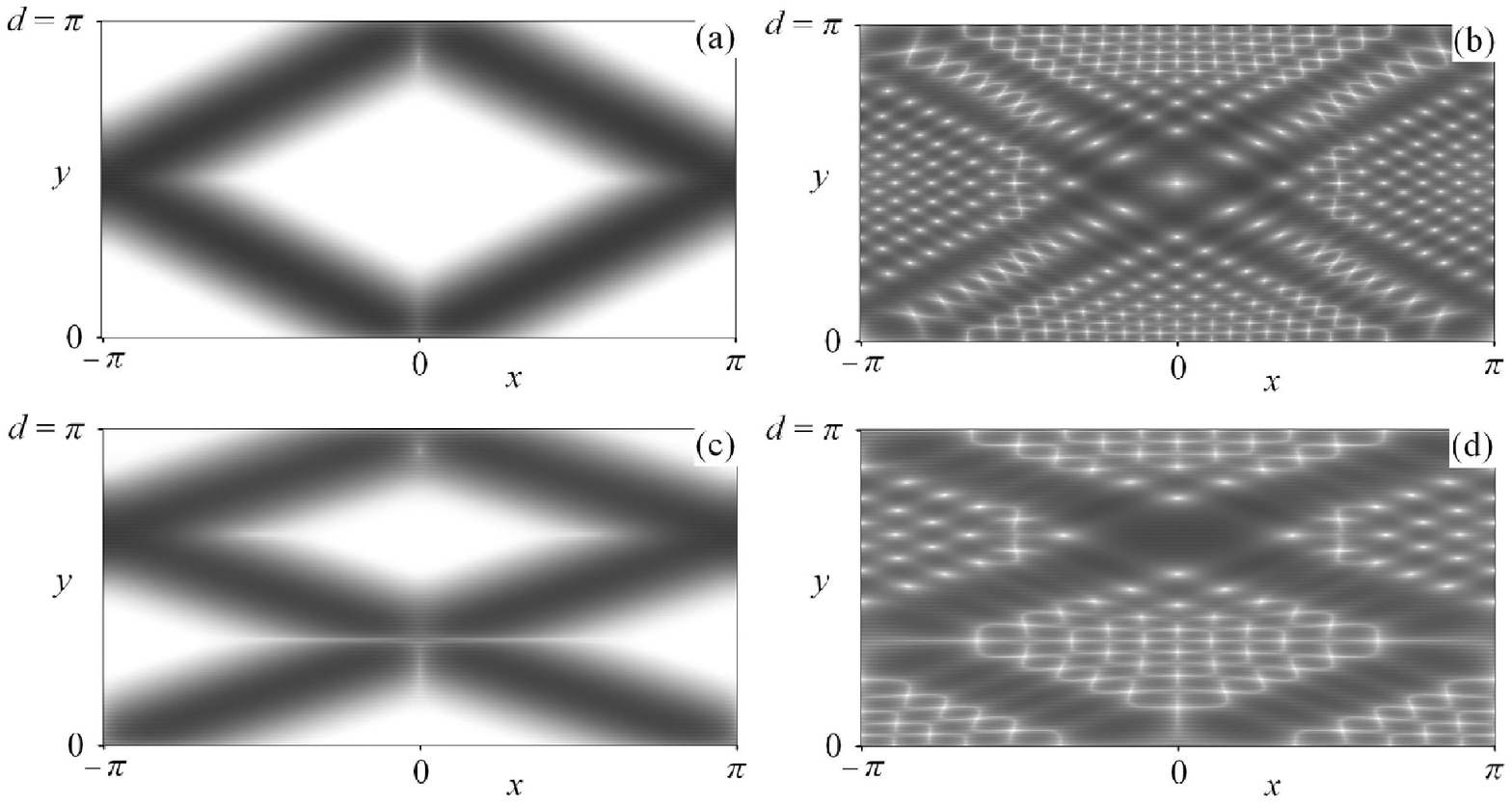}
 \caption{Examples of eigenstates probabilities $|\psi(x,y)|^2$
 (for resonance centre and for near-separatrix state)
 in grayscale for the same model parameters, but for the resonances
 $\eta=1/2$ (a), (b) and $\eta=1/3$ (c), (d).}
 \label{distr3}
 \end{center}
\end{figure}

Using the same technique we have calculated also quantum states
for the resonances $\eta=1/2$ and $\eta=1/3$. These resonances are
somewhat narrow than $\eta=1$ (see Fig.~\ref{fclass}({\it b}))
and the number of under-separatrix
levels are smaller. As an example Fig.~\ref{distr3} illustrates the
probability distributions $|\psi(x,y)|^2$ for the resonance centers and
near-separatrix states for $\eta=1/2$ ((a) and (b)) and $\eta=1/3$
((c) and (d)) correspondingly. It is clear that their structures are
also reflect the character of correspondent classical trajectories
at these resonances.

\section{Evolution operator}   
In this section we consider the dynamics of a charged particle in the
rippled channel in the presence of the time-dependent electric
field described by the potential $V(y,t)= - f_0 y (\cos \Omega_1 t
+ \cos \Omega_2 t)$. We assume that the frequencies $\Omega_1$ and
$ \Omega_2 $  are chosen to fulfill the condition $\omega_{n_0} =
\left( \Omega_1 + \Omega_2 \right)/2$ in order to provide equal
driving forces for a particle inside the stochastic layer of the
separatrix at the main coupling resonance. Specifically, we take,
$\omega_{n_0}=400, \,\, \Omega_1 = 350, \,\, \Omega_2 =450$,
therefore, the period $T$ of the perturbation is, $T=7\cdot
2\pi/\Omega_1 = 9 \cdot 2\pi/ \Omega_2 \approx 0.126$.

\begin{figure}[!tb]
 \begin{center}
 \includegraphics[width=4.0in]{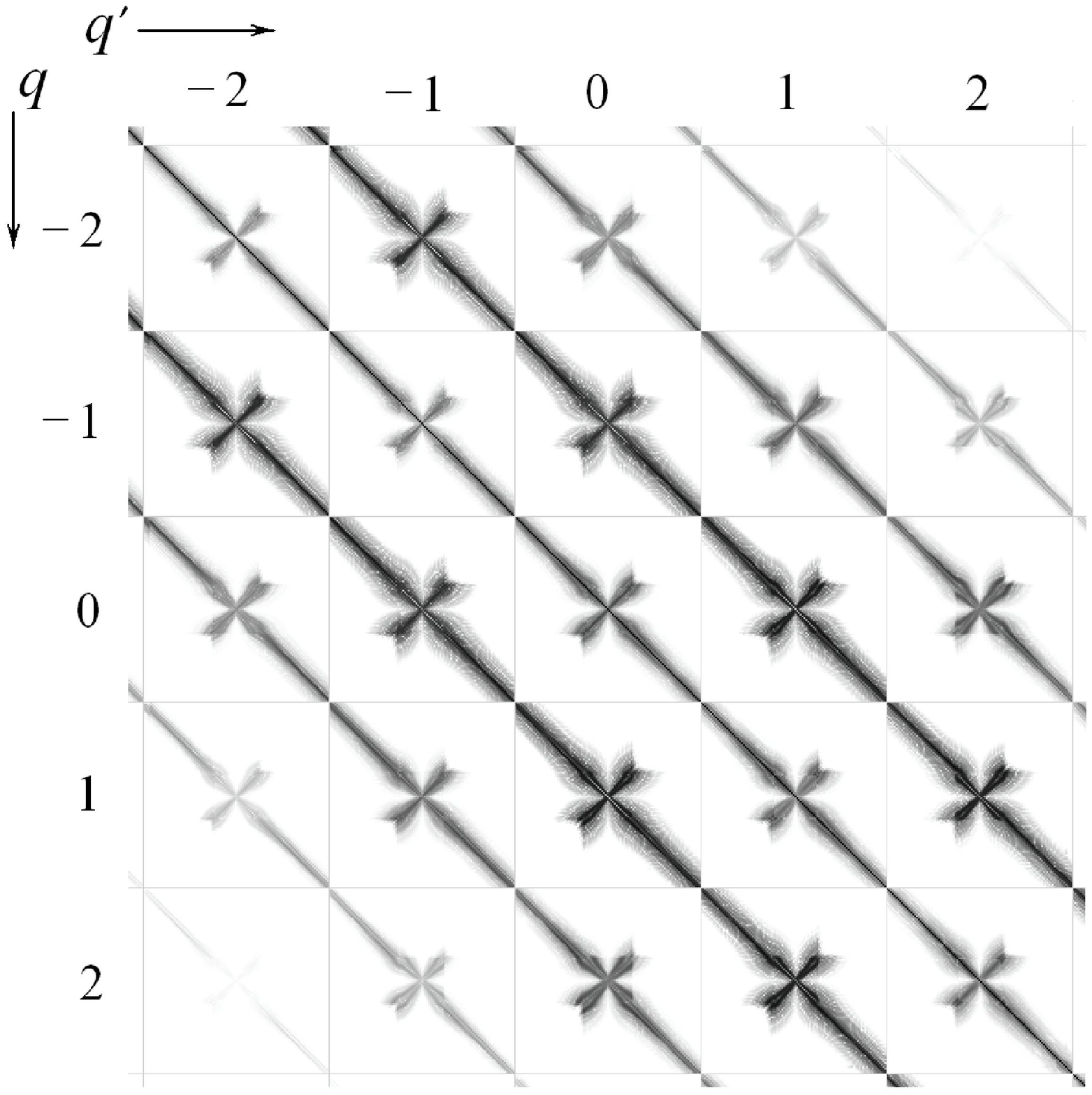}
 \end{center}
\caption{The distribution of the matrix elements modulus
   $|U_{q,s,q',s'}(T)|$ for the same parameters as in
   Fig.~\ref{distr1} and $f_0 = 5$. There are $101\times 101$
   elements in each block.}
  \label{evstr}
\end{figure}

Since the total Hamiltonian is periodic in time, one can write the
solution of the non-stationary Schr\"odinger equation as
$\psi_Q(x,y,t) = \exp \left( -i\epsilon_Q t\right) u_Q (x,y,t)$,
where $u_Q (x,y,t+T)=u_Q (x,y,t)$ is the quasienergy (QE) function
and $\epsilon_Q$ is the quasienergy. It can be shown that the QE
functions are the eigenfunctions of the evolution operator
$\hat U(T)$ of the system for one period of the perturbation. The
procedure to determine this operator was described in details in
Ref.~\cite{DIM02}. The matrix elements $U_{q,s,q',s'}(T)$ of the
evolution operator can be calculated by means of the numerical
solution of the non-stationary Schr\"odinger equation for
different initial states. The eigenvectors $A^Q_{q,s}$ (in the
representation of stationary problem with eigenfunctions~(\ref{psi}))
and eigenvalues $\exp{(-i\varepsilon_QT)}$ can be
found numerically by a direct diagonalization of the evolution
operator matrix. After that, the evolution matrix $U_{q,s,q',s'}(NT)$
for $N$~pe\-riods can be written as
\begin{eqnarray}
  \label{ev_oper}
  U_{q,s;q',s'}(NT)=\sum_{Q} A^Q_{q,s} {A^{Q^{\,*}}_{q',s'}}
  \exp{(-i\varepsilon_QNT)}.
\end{eqnarray}
Therefore, the evolution of any initial state can be computed by
making use of the evolution matrix,
\begin{eqnarray}
  \label{CC}
  C_{q,s}(NT)=\sum _{q',s'} U_{q,s,q',s'}(NT)C_{q',s'}(0).
\end{eqnarray}

In order to illustrate the structure of matrix $U_{q,s,q',s'}(T)$
we plot the matrix elements modules in Fig.~\ref{evstr}. As one
can see, this matrix has global block structure, with a cross-like
structure at the centers of blocks. When the evolution operator
acts on the initial state $C_{q,s}=\delta_{q,q_0}\delta_{s,s_0}$,
the resultant state will coincide with the $(q_0,s_0)$-column of
the matrix. The edges of \lq\lq{}crosses\rq\rq{}
at the block's centers where
matrix elements are relatively large, corresponds to a transition
between separatrix states. As a result, the transition between
such states of neighbor groups (along the coupling resonance) is
much stronger than those between other states. Note that the
separatrix states are responsible for the quantum Arnol'd
diffusion.

\begin{figure}[!tb]
 \begin{center}
 \includegraphics[width=4.0in]{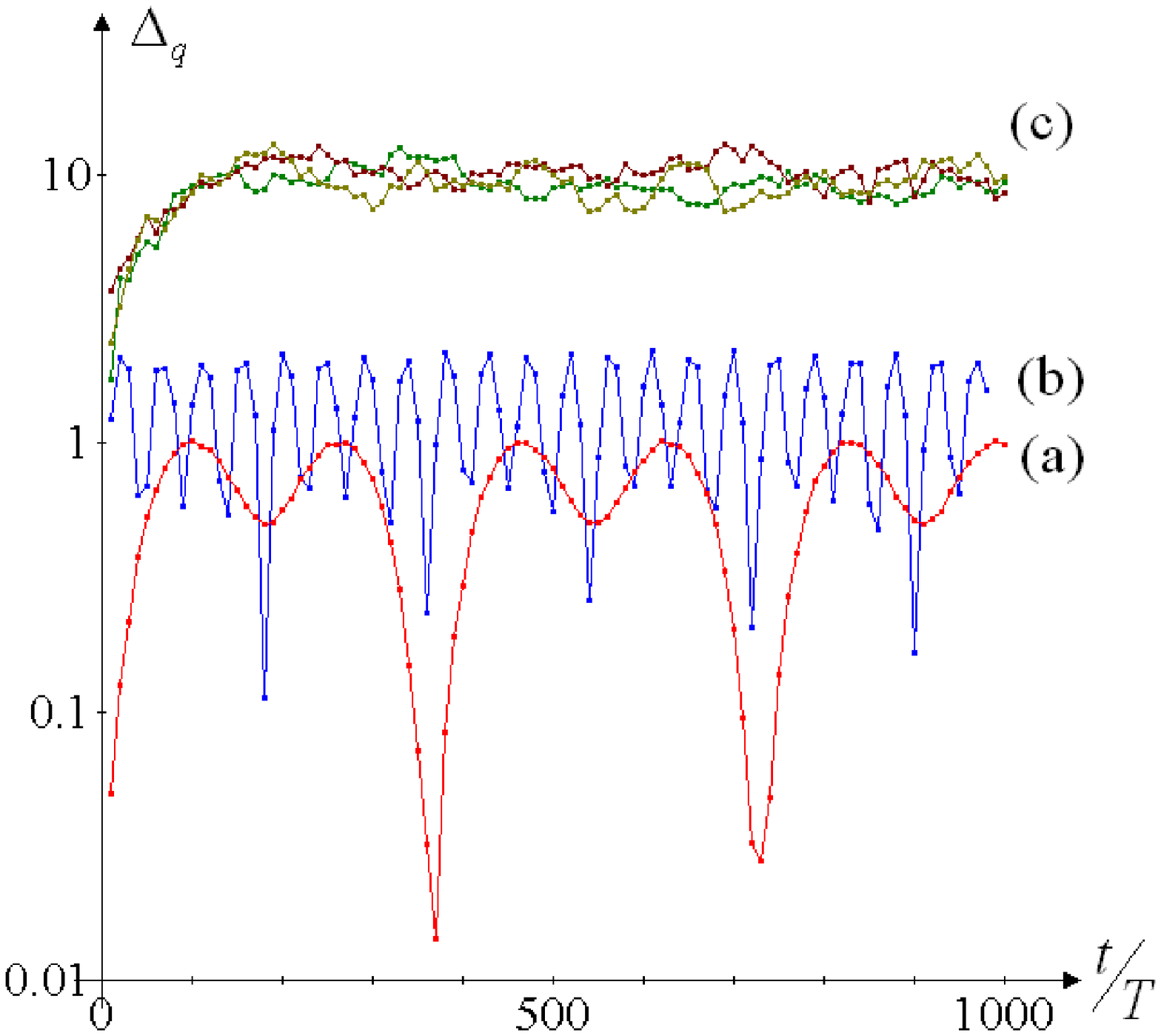}
 \end{center}
 \caption{Time dependence of the variance $\Delta_q$ for different
  initial states at the coupling resonance for the group with $q = 0$:
  (a) lowest level with $s = 0$, (b) above-separatrix level
  with $s = +45$, (c) near-separatrix levels with $s = -21, +22, -23$. Here
  $a = 0.01 $ and $f_0 = 10$. }
  \label{var}
\end{figure}

Our goal is to analyze the dynamics of a particle, initially
placed inside the separatrix under the condition that the coupling
and two driving resonances do not overlap. In Fig.~\ref{var}
typical dependencies of the variance
$\Delta_q=\overline{\left(\Delta \bar H \right)^2}/\omega^2_{n_0}$
of the normalized energy are shown versus the time measured in the
number~$N$ of periods of the external perturbation, for different
initial conditions. Here the quantity $\Delta_q$ is defined as the
variance of a wave packet in the $q$-space, $\Delta_q= \sum_q
(q-\bar q)^2\sum_s |C_{q,s}|^2$, where $\bar q = \sum_q q \sum_s
|C_{q,s}|^2$.

The data presented at Fig.~\ref{var} clearly demonstrate
a different character of the evolution of the system
in dependence on initial state. For the
state taken from the center of the coupling resonance, as well as
from above the separatrix, the variance oscillates in time, in
contrast with the state taken from inside the separatrix. In the
latter case, after a short time the variance of the energy
increases linearly in time, thus manifesting a diffusion-like
spread of the wave packet.

In order to characterize the speed of the diffusion, we have
calculated classical diffusion coefficients $D_1$, $D_{1/2}$ and
$D_{1/3}$ for the resonances $\eta=1$, $\eta=1/2$ and $\eta=1/3$,
correspondingly:
  \begin{eqnarray}
    D=\overline{\Biggl(
    \frac{(\bar E_{i+1}-\bar E_i)^2}{(\Delta t_i+\Delta t_{i+1})/2}
    \Biggr)}.
   \label{D}
  \end{eqnarray}
To suppress large fluctuations of the energy in time and to reveal
a stochastic character of motion, the averaging in Eq.~(\ref{D})
was performed in two stages. Here $\bar E_i$ is the average value
of the particle energy over the time $\Delta t_i$ corresponding to
$10^3$ collisions with the rippled wall. The second average has
been done in the following way. Having the mean value $\bar E_i$
in each interval $\Delta t_i$, the difference $(\bar E_{i+1}-\bar
E_i)$ between adjacent intervals was computed, and after that the
expression in brackets in Eq.~(\ref{D}) was averaged over all
these differences.

\begin{figure}[!tb]
 \begin{center}
 \includegraphics[width=4.0in]{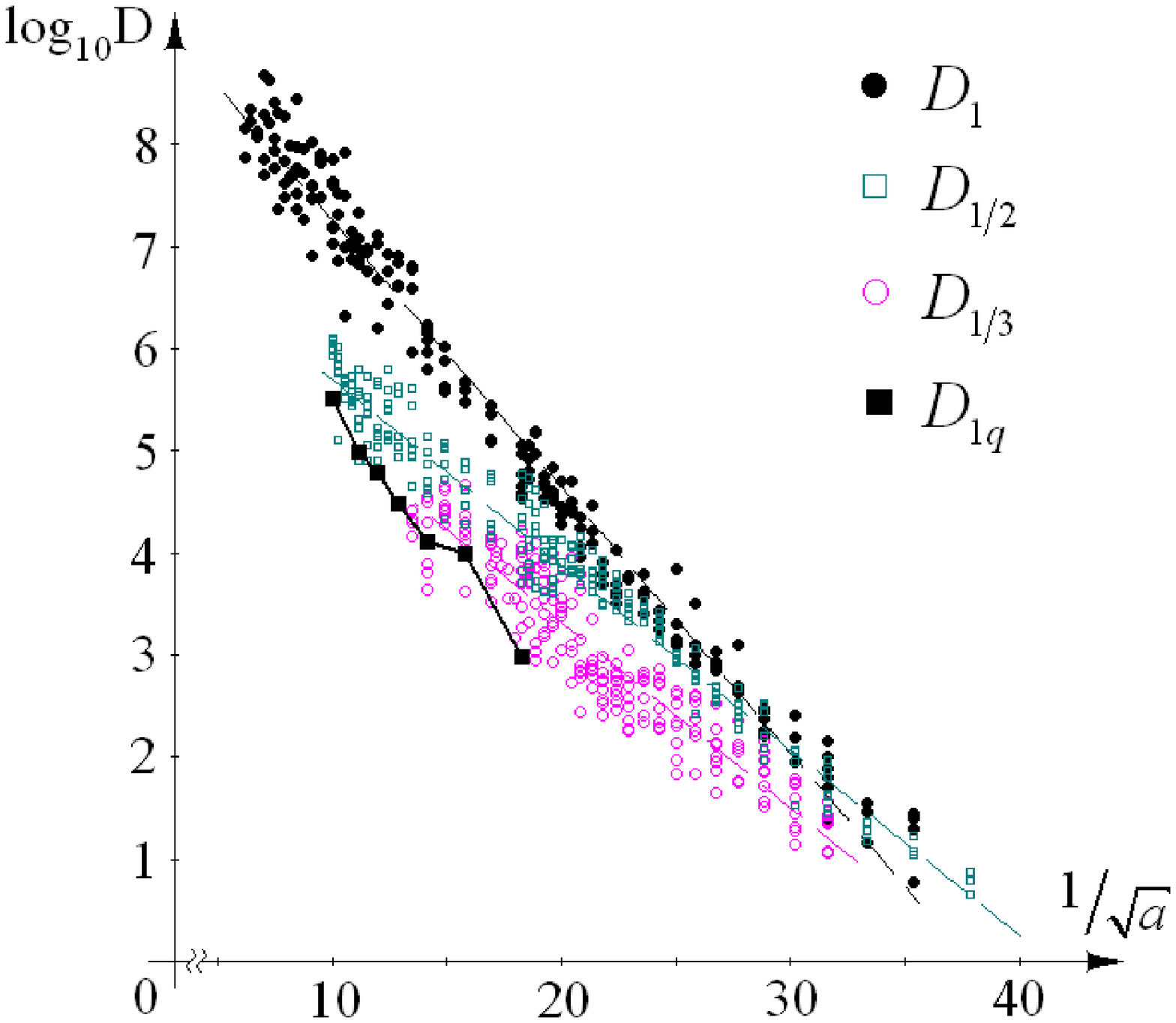}
 \end{center}
\caption{Classical, $D_1$, $D_{1/2}$ and $D_{1/3}$, versus
quantum, $D_{1q}$, diffusion coefficients in the dependence on the
amplitude $a$ of the rippled profile.}
  \label{coeff}
\end{figure}

The quantum diffusion coefficient $D_{1q}$ was calculated only for
the resonance $\eta=1$, see Fig.~\ref{coeff}. It was found that
the quantum Arnol'd diffusion roughly corresponds to the classical
one. However, the data clearly indicate that the quantum diffusion
is systematically weaker than the classical Arnol'd diffusion. In
all cases the ratio between the frequencies $\Omega_1$ and
$\Omega_2$ and $\omega_{m_0}$ of the external field was the same,
$\Omega_1:\omega_{m_0}:\Omega_2=7:8:9$.

One should stress that the quantum Arnol'd diffusion takes place
only in the case when the number $M_s$ of energy stationary states
in the separatrix layer is relatively large. For the first time,
this point was noted by Shuryak~\cite{S76} who studied the
quantum-classical correspondence for nonlinear resonances. In this
connection we have estimated the number of the energy states that
occupy the separatrix layer. We have found that for $a = 0.01$ the
number $M_s$ of stationary states inside the separatrix chaotic
layer is more than $10$, therefore, one can speak about a kind of
quantum chaos in this region. On the other hand, with a decrease
of the goffer amplitude, the number $M_s$ decreases and for
$1/\sqrt{a}\approx 20$ it is of the order of one. For this reason
the last right point in Fig.~\ref{coeff} corresponds to the
situation when the chaotic motion along the coupling resonance is
completely suppressed by coherent quantum effects. This effect is
known as the \lq\lq{}Shuryak border\rq\rq~\cite{S76}) that has to
be taken into account when considering the conditions for the
onset of quantum chaos.

Since the diffusive motion along the coupling resonance is
effectively one-di\-men\-sional, one can naturally expect an
Anderson-like localization. Indeed, the va\-ri\-ance of the QE
eigenstates of the evolution operator is finite in the $q$-space.
This means that eigenstates are localized, and the wave packet
dynamics along the separatrix layer has to reveal the saturation
of the diffusion. More specifically, one expects that the linear
increase of the variance of the energy ceases after some
characteristic time.

In order to observe the dynamical localization in our model (along
the coupling resonance and inside the separatrix layer), one needs
to study long-time dynamics of wave packets. Our numerical
analysis for large times (see curves (c) on Fig.~\ref{var}) has
confirmed that after some time $t \sim 200 T$, the diffusion-like
evolution stops for different values of the amplitude $a$. For
larger times, the variance $\Delta_q$ starts to oscillate around
its mean value.

This effect, known as the {\it dynamical localization}, has been
discovered in~\cite{CCIF79,CIS81} for the kicked rotor, and was
studied later in different physical models (see, for example,
Ref.~\cite{LL92} and references therein). One should note that the
dynamical localization is, in principle, different from the
Anderson localization, since the latter occurs for models with
random potentials. In contrast, the dynamical localization happens
in dynamical (without any randomness) systems, and is due to the
interplay between (week) classical diffusion and (strong) quantum
effects.

\section{Summary}

Let us now summarize our main results. First, we have studied
stationary quantum states for a particle moving inside a 2D
rippled billiard. Main attention was paid to the states
corresponding to the coupling resonance~$\eta=1$. We have numerically
analyzed the structure of these states, by plotting their density
distributions~$|c^k_{r,p}|^2$ in the unperturbed basis
determined by the model with flat boundaries.
We have found also that the patterns in $(x,y)$-space of the
states corresponding to the inside-resonance region can be
associated with the classical orbits. Also, it was shown that the
states corresponding to the separatrix region, have a global
structure similar to the classical orbits as well. It was found
that the width of such states in the unperturbed basis is much
larger than of those belonging to the inside-resonance region.

In the presence of the external two-frequency time-dependent
electric field, the main interest is related to the dynamics of
wave packets along the narrow stochastic regions of the
corresponding classical system. Extensive numerical simulation
demonstrate a kind of weak diffusion in the quantum model, that
can be associated with the classical Arnol'd diffusion along the
coupling resonance. We have found that the dependence of the
diffusion coefficient on model parameters roughly follows the
classical dependence. However, the quantum diffusion is
systematically slower than the classical one. This fact manifests
the influence of quantum effects.

It should be stressed that the quantum Arnol'd diffusion occurs in
a deep se\-mi\-clas\-si\-cal region, specifically for the case when
the number $M_s$ of chaotic eigenstates inside the classical
stochastic layer is sufficiently large (of the order of 10 or
larger). With a decrease of the amplitude of ripple, the diffusion
coefficient strongly decreases, and for $M_s \leq 1$ the diffusion
disappears. Therefore, one can see how quantum effects destroy the
diffusive dynamics of the wave packets.

Another manifestation of quantum effects is the dynamical
localization that persists even for large $M_s$. Specifically, we
have observed that the quantum diffusion occurs only for finite
times. On a larger time scale the diffusion ceases and after some
characteristic time it terminates. This effect is similar to that
discovered in the kicked rotor model~\cite{CCIF79}, and found
later in other physical systems (see, for example,
Ref.~\cite{LL92} and references therein). In our case the
dynamical localization arises for a weak chaos that occurs inside
the separatrix layer, in contrast to previous models with a strong
(global) chaos in the classical description.

The effects discussed in this communication can be observed
experimentally, e.g. in semi-metal structures. Apart from the
experiments with a 2D electron gas in GaAs/AlGaAs
heterojunctions~\cite{DMH03,H99}, one can consider the semi-metal
rippled channel with a large number of discrete quantum levels in
a deep semiclassical regime. For the observation of the dynamical
localization in a response to electromagnetic radiation the
channel, one can take $d=1\,\mu$m for the width, and $l=2\,\mu$m
for the goffer period. Then, the dimensionless value $a=0.01$
corresponds to the goffer amplitude of the order 3.2~nm, i.e. to
few monoatomic levels. Correspondingly, for the effective electron
mass $m \simeq 0.1m_{\rm e}$ the level number $n_0=400$ has the energy
$E_{n_0}\simeq 0.61$~eV, and the resonance frequency for the
transitions between nearest states is $\omega_{n_0}/2\pi\simeq
740$~GHz. In these units the period of the external field is equal
to $1.1\times 10^{-11}$~s, and the dimensionless value of the
perturbation $f_0=10$ corresponds to an electric field $E \simeq
0.24$~V/cm. It should be noted that for above parameters the
diffusion saturation time, corresponding to approximately 200
periods, is of the order of $2.2$~ns. It is clear that the
electron scattering time has to be much larger. The last condition
can be realized in semi-metal Bi or Sb structures, where the mean
free path is of the order of 1~mm and the scattering time is about
10~ns for the temperature $\sim 1$~K.

\section*{Acknowledgements}
This work was supported by RFBR Grant no.~06-02-17189 and by the
program \lq\lq Development of scientific potential of high
school\rq\rq{} of Russian Ministry of Education and Science
(project no.~RNP.2.1.1.2363) and partially by the CONACYT
(M\'exico) grant No~43730. A.I.M. acknowledges the support of RFBR
Grant no.~06-02-16561.


\end{document}